\documentclass[conference,compsocconf]{IEEEtran}

\usepackage{booktabs} % For formal tables
\usepackage{mathptmx} % This is Times font
\usepackage{fancyhdr}
\usepackage[normalem]{ulem}
\usepackage[hyphens]{url}
\usepackage{hyperref}
\usepackage[caption=false,font=footnotesize]{subfig}
\usepackage{amssymb}
\usepackage{stfloats}
\usepackage{graphicx}

\begin{document}
%
% paper title
% can use linebreaks \\ within to get better formatting as desired
\title{Medusa: A Scalable Interconnect for Many-Port DNN Accelerators and Wide DRAM Controller Interfaces}

% author names and affiliations
% use a multiple column layout for up to two different
% affiliations

\author{
\IEEEauthorblockN{Yongming Shen}
\IEEEauthorblockA{Stony Brook University\\
yoshen@cs.stonybrook.edu}
\and
\IEEEauthorblockN{Tianchu Ji}
\IEEEauthorblockA{Stony Brook University\\
tianchu.ji@stonybrook.edu}
\and
\IEEEauthorblockN{Michael Ferdman}
\IEEEauthorblockA{Stony Brook University\\
mferdman@cs.stonybrook.edu}
\and
\IEEEauthorblockN{Peter Milder}
\IEEEauthorblockA{Stony Brook University\\
peter.milder@stonybrook.edu}}

% make the title area
\maketitle

% two following lines to enable page numbers
%\thispagestyle{plain}
%\pagestyle{plain}

%\thispagestyle{firstpage}
%\pagestyle{plain}

\begin{abstract}
To cope with the increasing demand and computational intensity of deep neural networks (DNNs), industry and academia have turned to accelerator technologies.
In particular, FPGAs have been shown to provide a good balance between performance and energy efficiency for accelerating DNNs.
While significant research has focused on how to build efficient layer processors,
the computational building blocks of DNN accelerators,
relatively little attention has been paid to the on-chip interconnects that sit between the layer processors and the FPGA's DRAM controller.

We observe a disparity between DNN accelerator interfaces,
which tend to comprise many narrow ports, and FPGA DRAM controller interfaces, which tend to be wide buses.
This mismatch causes traditional interconnects to consume significant FPGA resources.
To address this problem, we designed Medusa: an optimized FPGA memory interconnect which transposes data in the interconnect fabric, tailoring the interconnect to the needs of DNN layer processors.
Compared to a traditional FPGA interconnect, our design can reduce LUT and FF use by 4.7x and 6.0x, and improves frequency by 1.8x.
\end{abstract}

\section{Introduction}
\label{sec:introduction}

Deep neural networks (DNNs)~\cite{simonyan2014very,szegedy2015googlenet,krizhevsky2012imagenet,resnet} are used to solve challenging machine learning problems.
However, CPUs are failing to meet the high computational demand of DNNs.
GPUs provide sufficient performance, but are limited by their high power consumption.
In contrast, research has shown that FPGAs strike a good balance between performance and energy efficiency for accelerating DNNs.

A DNN comprises a pipeline of computing layers (3D convolution, sub-sampling, nonlinear activation, etc.).
Correspondingly, an FPGA-based DNN accelerator comprises one or more layer processors,
where each is specialized for computing one or more layers of the target DNN~\cite{li2016high,sharma2016dnnweaver,zhang2015optimizing,shen2017multiclp,shen2017escher}.
For large DNNs, DRAM is needed to store DNN parameters and layer inputs and outputs.
Prior work has shown that DNN computation is highly bandwidth intensive~\cite{shen2017escher,alwani2016fused}.
It is thus essential for the layer processors to fully utilize the available DRAM bandwidth.
However, there exists a mismatch between the interface of an FPGA DRAM controller and the layer processors.
The nature of FPGAs tends to restrict the frequency of layer processors,
which results in the DRAM controller using a wide interface to expose the full DRAM bandwidth to the layer processors (512-bits for a single DDR3 channel).
On the other hand, many state-of-the-art FPGA-based DNN accelerators~\cite{shen2017multiclp,shen2017escher} assume the availability of many narrow read and write ports (8 or 16 bits), each with independent DRAM access.
This is because narrow ports offer the most flexibility in optimizing the layer processors for the target DNN~\cite{shen2017multiclp}.
As such, a memory interconnect must be used to multiplex the wide DRAM controller interface to a large number of narrow read and write ports, while maintaining maximum bandwidth efficiency.

A memory interconnect performs data transfer as well as request arbitration.
The challenge of multiplexing a wide DRAM controller interface lies in \emph{data transfer}, which will be our focus.
Mainstream designs of memory interconnects~\cite{XilinxAXI,AlteraQsys} use a 1-to-$N$ crossbar to multiplex the wide DRAM controller interface to $N$ narrower ports.
The crossbar needs to have the same width as the DRAM controller to ensure that the memory bandwidth is fully utilized.
Each of the $N$ endpoints of the crossbar must then connect to a FIFO to buffer burst transfers and a data-width converter to present a narrow port to the DNN accelerator.
While straightforward, such designs are severely over-provisioned:
the wide crossbar allows the full DRAM bandwidth to be directed to any narrow port on any cycle, but each narrow port only uses a fraction of the full bandwidth.
This excessive flexibility of the interconnect consumes significant logic and wiring resources that can otherwise be used by the DNN accelerator.

To overcome this over-provisioning, a memory interconnect should be optimized to take advantage of the data transfer characteristics of DNN layer processors.
In this regard, we make two critical observations.
First, the narrow ports used by layer processors are all of the same width, and are all expected to be able to supply one word per cycle.
This means that DRAM bandwidth should be statically and evenly partitioned across the narrow ports.
Second, a layer processor knows its access pattern and can perform perfect prefetch for future data access,
which means that a moderate latency increase in the memory interconnect will not affect system performance.

Based on our observations, we designed Medusa, a resource-efficient, performant, and scalable memory interconnect.
In our design, the crossbar, FIFOs, and data-width converters are replaced with a transposition unit.
Within the transposition unit, a shifter replaces the crossbar and data-width converters, resulting in significant logic simplification.
Moreover, instead of a shallow FIFO per port, the transposition unit uses a deep shared buffer.
This allows BRAMs to be efficiently used for buffering, freeing up LUTs and wires for other uses.
Importantly, with only a minor \emph{constant} latency increase, Medusa guarantees the same data transfer characteristics as the traditional interconnects, and can be used as a drop-in replacement without changing the layer processor or memory request arbiter design.

Compared to a traditional interconnect,
Medusa multiplexes a 512-bit DRAM controller interface across 32 16-bit read ports and 32 16-bit write ports
using 4.7x and 6.0x fewer LUTs and FFs, while also improving frequency by 1.8x.
For a 1024-bit DRAM controller interface, Medusa runs at 225MHz, while routing congestion limits traditional designs to under 25MHz.

\section{Traditional Memory Interconnects}
\label{sec:baseline}

In this section, we present a baseline memory interconnect which is representative of existing designs~\cite{XilinxAXI,AlteraQsys}, and we qualitatively discuss its scalability challenges.

\subsection {Baseline Data Transfer Logic}

The baseline interconnect's data transfer logic has two parts:
one for memory read, and the other for memory write.

\subsubsection{Baseline Memory Read Data Transfer}

\begin{figure}
\centering
\includegraphics[width=\columnwidth]{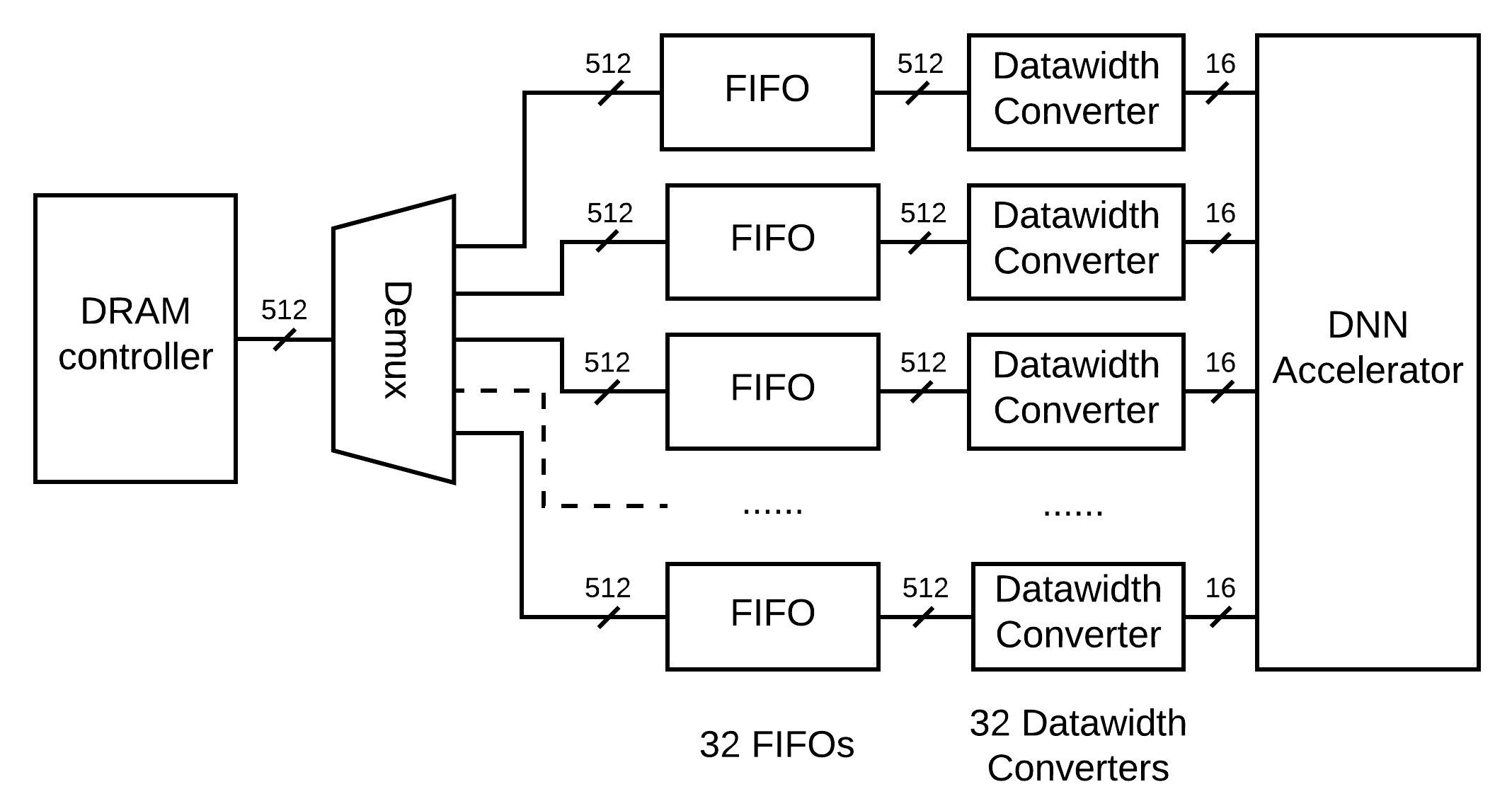}
\caption{The baseline memory read data transfer network.}
\label{fig:baseline_read_data_network}
\end{figure}

Figure~\ref{fig:baseline_read_data_network} shows the baseline memory read data transfer network. In this example, 32 16-bit accelerator read ports share access to one 512-bit DRAM controller interface.
The design uses a 1-to-$N$ demux to route input data from the memory controller to $N$ FIFOs, where each FIFO has the same width as the memory interface.
This means that the demux can accept a new input from the memory controller on every cycle, allowing the maximum memory bandwidth to be consumed. 
Each FIFO is provisioned to be large enough to hold the largest burst that a narrow read port can request, so that burst transfer to a single narrow port does not create back pressure.
The output of each FIFO is connected to a data width converter, which converts data from the memory interface width to the narrow read port width.

\begin{figure}
\centering
\includegraphics[width=\columnwidth]{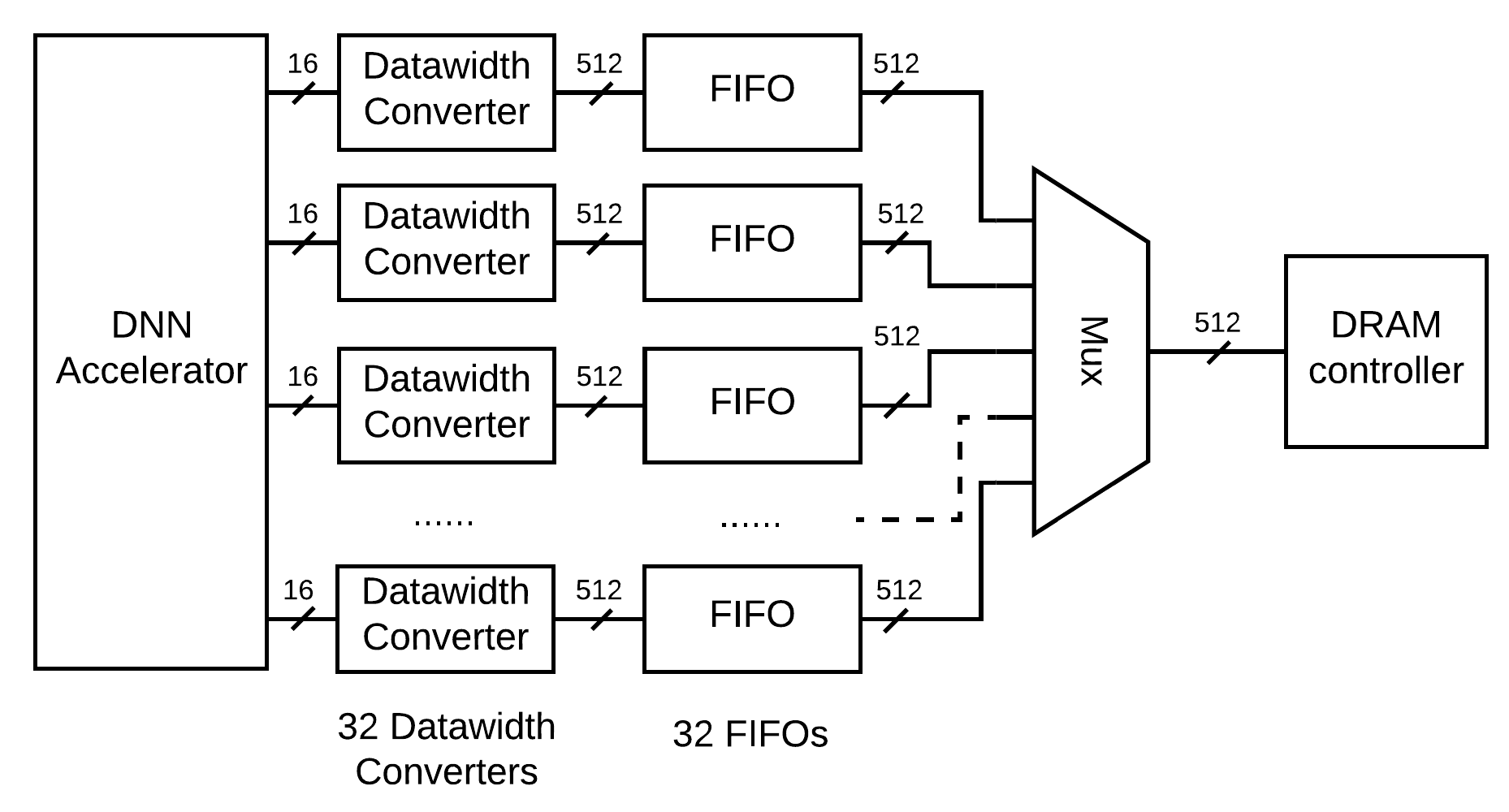}
\caption{The baseline memory write data transfer network.}
\label{fig:baseline_write_data_network}
\end{figure}

\subsubsection{Baseline Memory Write Data Transfer}

Figure~\ref{fig:baseline_write_data_network} shows the baseline memory write data transfer network, which is similar to the read data transfer network, except the data words flow in the opposite direction.
Each of the $N$ accelerator write ports feeds into a data width converter, then into a FIFO.
Each FIFO has the same width as the memory controller and can hold the maximum burst from a port.
On each cycle, an $N$-to-1 mux chooses the output from one of the FIFOs to write to the memory controller.
By using FIFOs to accumulate complete bursts of data, data from the same burst can be sent to the memory controller using the full bandwidth of the memory controller interface.

\subsection{Baseline Logic Complexity}

To better understand the baseline design, we perform an analysis of its logic complexity.
Given a memory controller interface data width with $W_{line}$ bits and a total of $N$ read (or write) ports on the accelerator(s), then each port must have $W_{acc}= W_{line}/N$ bits for the complete memory bandwidth to be utilized. 
For simplicity, this analysis assumes both $W_{line}$ and $N$ to be powers of 2.

For the baseline read data transfer network, the main logic resource use comes from the data width converters and can be measured in terms of the number of {2}-to-{1} single bit muxes.
Each converter needs to perform an {$N$}-to-{1} mux of width $W_{acc}$ (or a shift of $W_{acc}$ bits); either method will have a cost of $W_{acc}\times (N-1)$ {2}-to-{1} single bit muxes.
With $N$ read ports, the total cost is $W_{acc}\times (N-1) \times N$ = $W_{line}\times (N-1)$ {2}-to-1 muxes.
In other words, the complexity is $O(Bandwidth\times NumPorts)$.

For the baseline write data transfer network, most of the logic is used for the {$N$}-to-{1} mux of width $W_{line}$, so the total resource use complexity is $W_{line}\times (N-1)$,
equal to the read data transfer network.

\subsection {Baseline Scalability Problems}

Although the baseline presents a straightforward solution to allow full DRAM bandwidth usage and to eliminate any data switching conflicts among narrow memory ports, its wide demux and mux are over-provisioned in terms of their connectivity.
For example, the demux used in the read network has the ability to direct \emph{all} of the read bandwidth to any of the read ports on any cycle.
Such flexibility is useful in applications where the partitioning of memory bandwidth to read ports needs to change over time.
However, in the context of DNN accelerators, the memory read bandwidth is expected to simply be evenly divided among all the read ports~\cite{shen2017multiclp}.
As such, the extra flexibility of the wide demux only incurs wasted logic (the muxes of the data width converters) and wiring resources.
The write network incurs analogous resource waste.

Moreover, the combination of wide and shallow FIFOs leads to  inefficient use of FPGA resources.
Implementing the shallow FIFOs using BRAMs wastes BRAM capacity, while using LUTRAM consumes a large amount of logic.
Additionally, a large number of buses (as wide as the DRAM controller interface) is widely distributed within this design.
Handling wide buses introduces challenges with FPGA routing, greatly limiting the peak clock frequency when scaling to wider memory interfaces.

\section{Medusa: An Optimized Memory Interconnect}
\label{sec:design}

\begin{figure*}[t]
\centering
\subfloat[Medusa's memory read data transfer network \label{fig:topview_read}]{\includegraphics[width=0.7\textwidth]{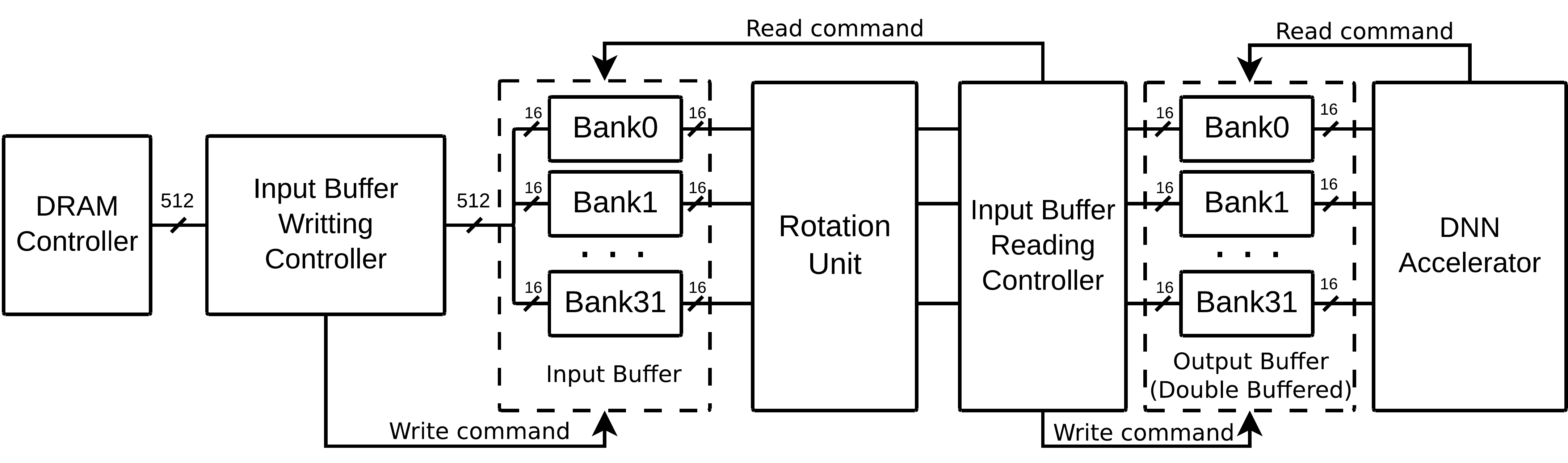}}\hfill \\
\subfloat[Medusa's memory write data transfer network \label{fig:topview_write}]{\includegraphics[width=0.7\textwidth]{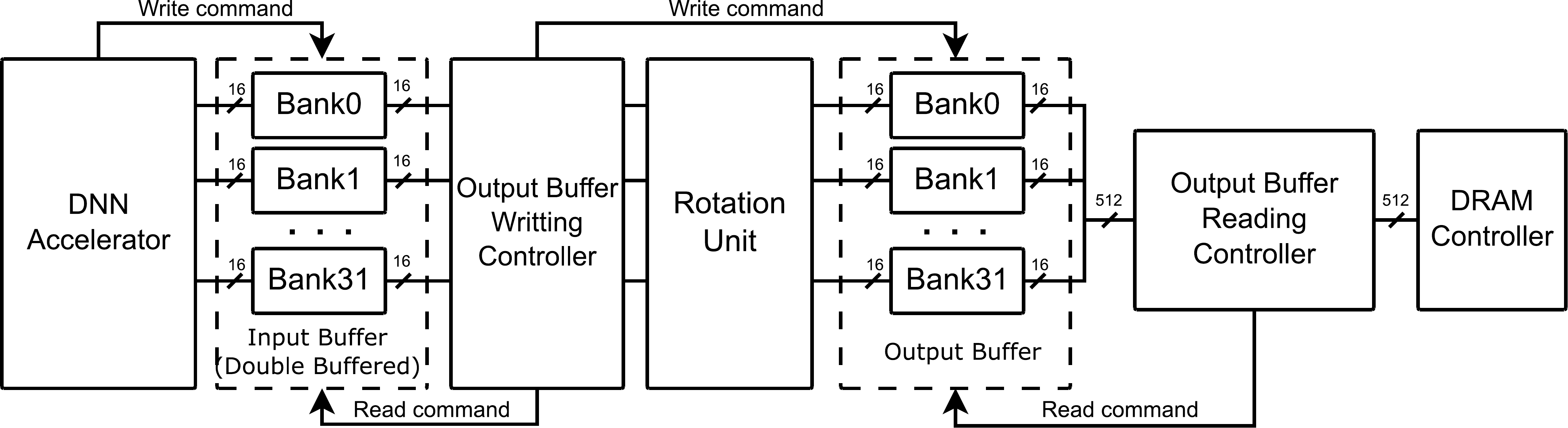}}\hfill
\caption{High level view of the memory interconnect Medusa. Controller modules keeps track of data and space availability in buffers. Buffers next to the DNN accelerator are double buffered. A line from the DRAM controller is 512-bit ($W_{line}$). Each port of the DNN accelerator is 16-bit wide ($W_{acc}=16$ bits).} \label{fig:topview}
\end{figure*}

We propose a scalable high performance memory interconnect which is based on data transposition.
Figures~\ref{fig:topview_read} and~\ref{fig:topview_write} provide a high-level overview of the interconnect.
Both memory read and write use two data buffers, a rotation unit, and control logic.

Our design evenly partitions the DRAM bandwidth to each port of the DNN accelerator by transposing data instead of routing it with wide demuxes and muxes, thus reducing FPGA resource and routing complexity,
without compromising DRAM bandwidth utilization.

\subsection{Bandwidth Partitioning Through Transposition}

Here we provide detailed descriptions of how transposition is used for memory read and write.
%For clarity, we ignore the impact of burst transfers and pipelining in this description.
%Support for burst transfers is presented in Section~\ref{ssec:burst}.

\subsubsection{Transposition for Memory Read}
\begin{figure*}[ht]
    \centering
    \subfloat[Cycle 0. \label{fig:detailed_read_transpose_exampleA}]{\includegraphics[width=0.32\textwidth]{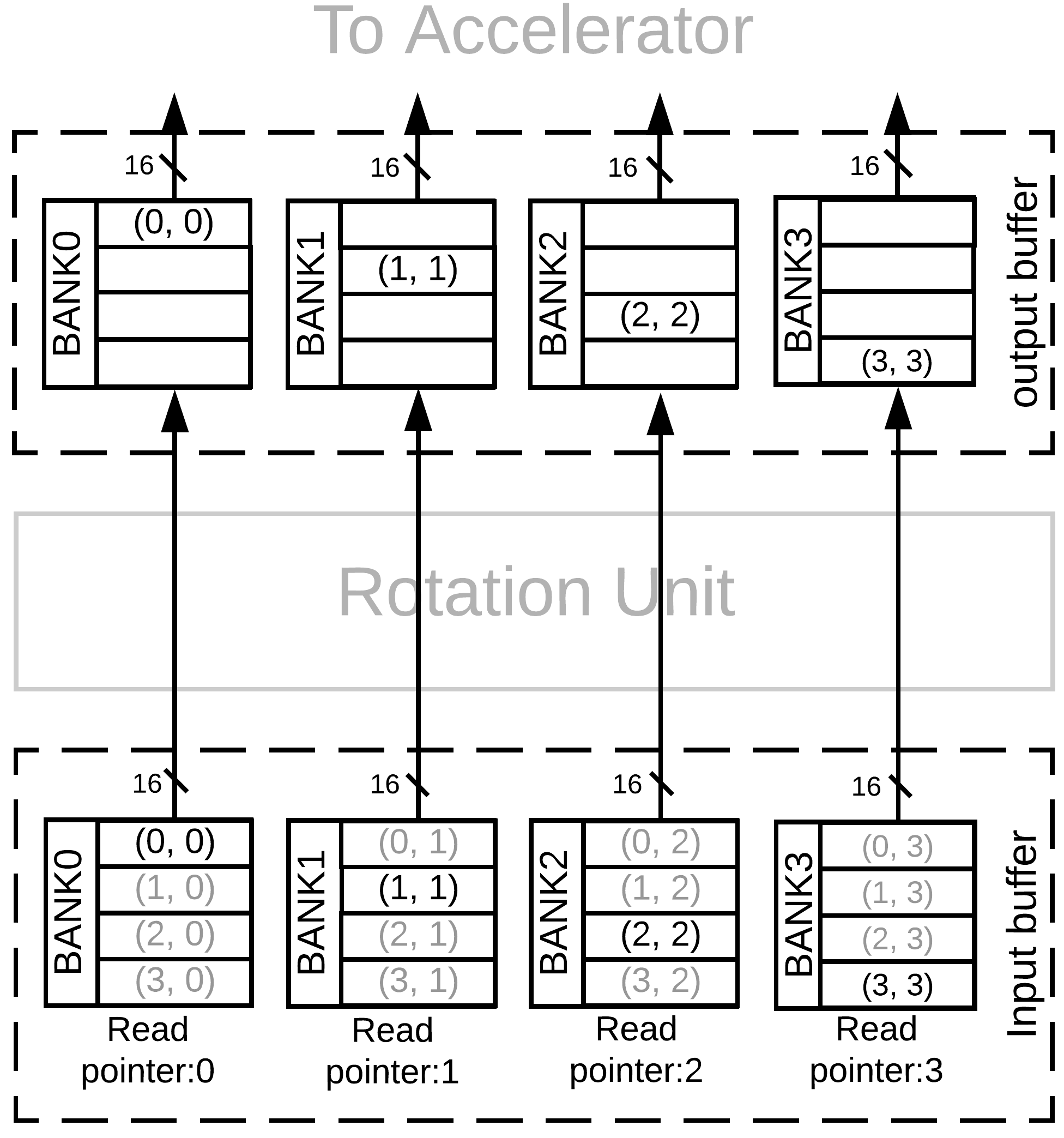}}\hfill
    \subfloat[Cycle 1. \label{fig:detailed_read_transpose_exampleB}]{\includegraphics[width=0.32\textwidth]{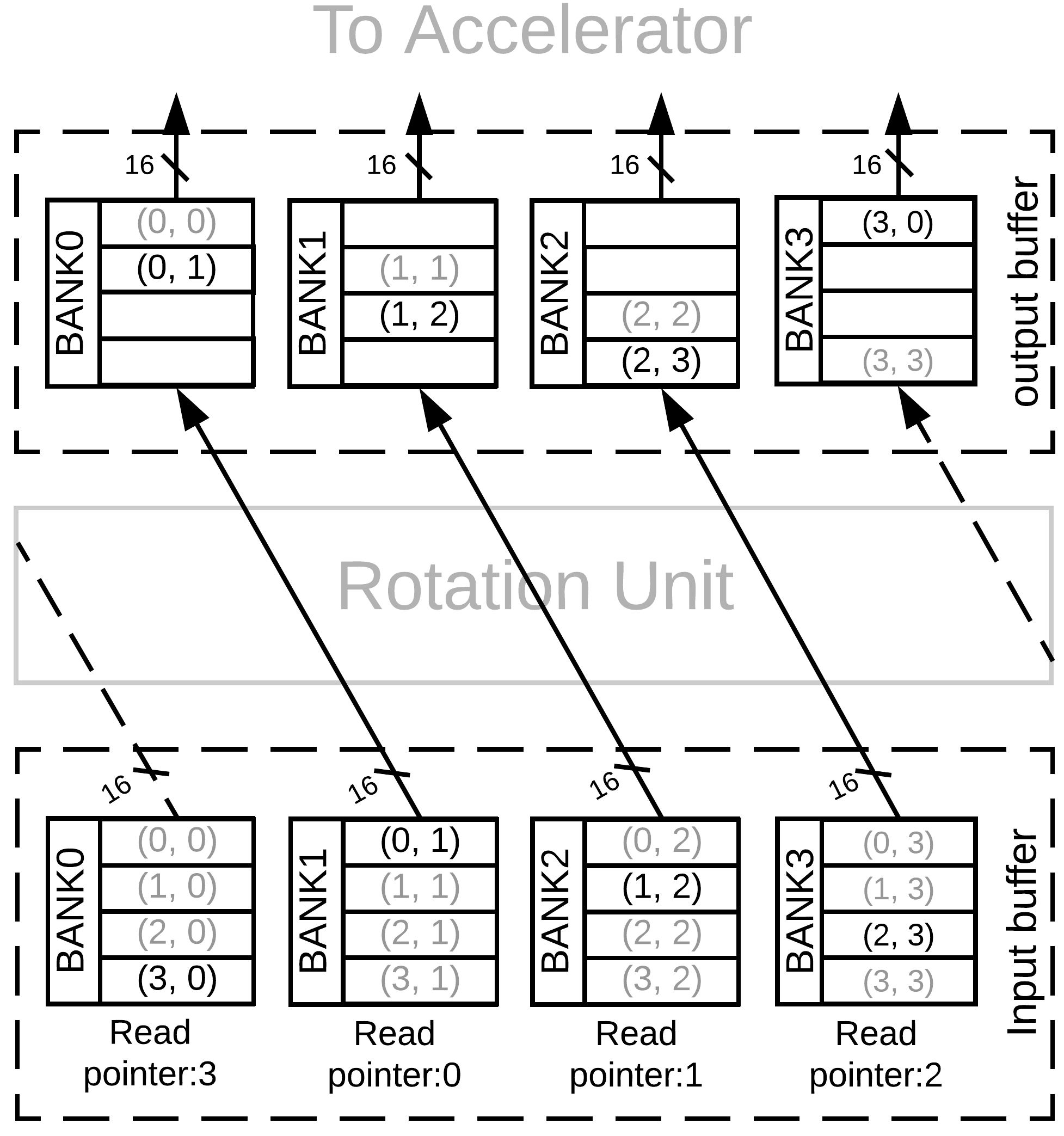}}\hfill
    \subfloat[Cycle 3 with transposition result. \label{fig:detailed_read_transpose_exampleC}]{\includegraphics[width=0.32\textwidth]{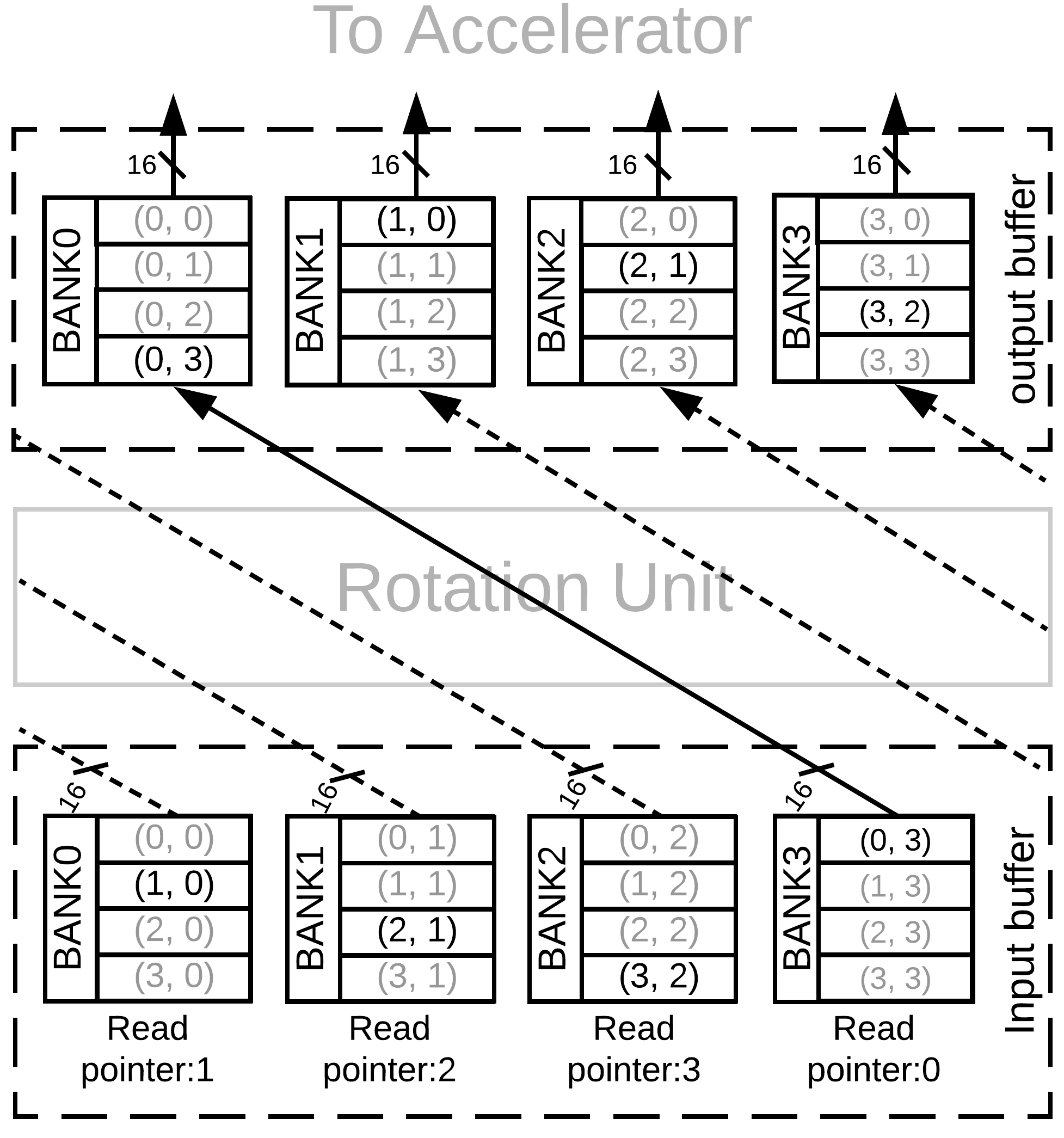}}\hfill
    \caption{A detailed transposition example for memory read.}
    \label{fig:detailed_read_transpose_example}
\end{figure*}

Figure~\ref{fig:detailed_read_transpose_example}
shows an example of transposition for memory read.
Each memory line is $W_{line} = 64$ bits,
each accelerator port is $W_{acc} = 16$ bits wide,
and $N=W_{line}/W_{acc} = 4$ accelerator ports are used.
We mark each data word with coordinates $(x,y)$, where $x$ represents the word's destination accelerator port, and $y$ is the word's index within its containing memory line.
Words in the same memory line are always destined to the same accelerator port, and are sent to the destination port in increasing index order.
Each $W_{line}$-bit memory line is stored across the input buffer banks (seen at the bottom of the figure).
Specifically, words that are destined to accelerator port $i$ are stored in address $i$ of each of the input buffer banks.

Transposition is performed by reading data words from the input buffer, rotating them, and storing them in appropriate locations in the output buffer.
First, at cycle $c$, words along the diagonal $(0, c\bmod N)$ to $(N-1, (c+N-1) \bmod N)$ are read.
For example, Figure~\ref{fig:detailed_read_transpose_exampleA} shows $c=0$, where words $(0,0), \dots, (3,3)$ are read, and  Figure~\ref{fig:detailed_read_transpose_exampleB} shows $c=1$, where words $(0,1), \dots, (3,0)$ are read.
The rotation unit then takes these $N$ words and rotates them to the left by $c \bmod N$ locations.
For example, Figure~\ref{fig:detailed_read_transpose_exampleC} shows that during $c=3$, the words are rotated 3 positions to the left.
Lastly, the output buffer stores the words into transposed locations: on cycle $c$, bank $i$ will store data into address $(i+c)\bmod N$.
The transposition completes in $N$ cycles, after which each accelerator port can read from its corresponding output buffer bank.

\subsubsection{Transposition for Memory Write}

Memory writes are performed similarly, but with data flowing in the opposite direction, as shown in Figure \ref{fig:topview_write}.
Each accelerator port writes data words into its own bank of the input buffer.
The interconnect then transposes input buffer banks to rows in the output buffer.

For both memory read and write, the interconnect is capable of processing one $W_{line}$-bit line per cycle, as all parts (the rotation unit, input buffer read/write, output buffer read/write) operate on $W_{line}$-bit data in parallel. Therefore the system can deliver the full bandwidth of the DRAM controller interface to the accelerator ports.
Furthermore, the bandwidth is evenly partitioned across the ports, matching the accelerator's requirements.

\subsection {Rotation Unit Design}
\label{ssec:rotation_unit}

\begin{figure}[ht]
\centering
\includegraphics[width=\columnwidth]{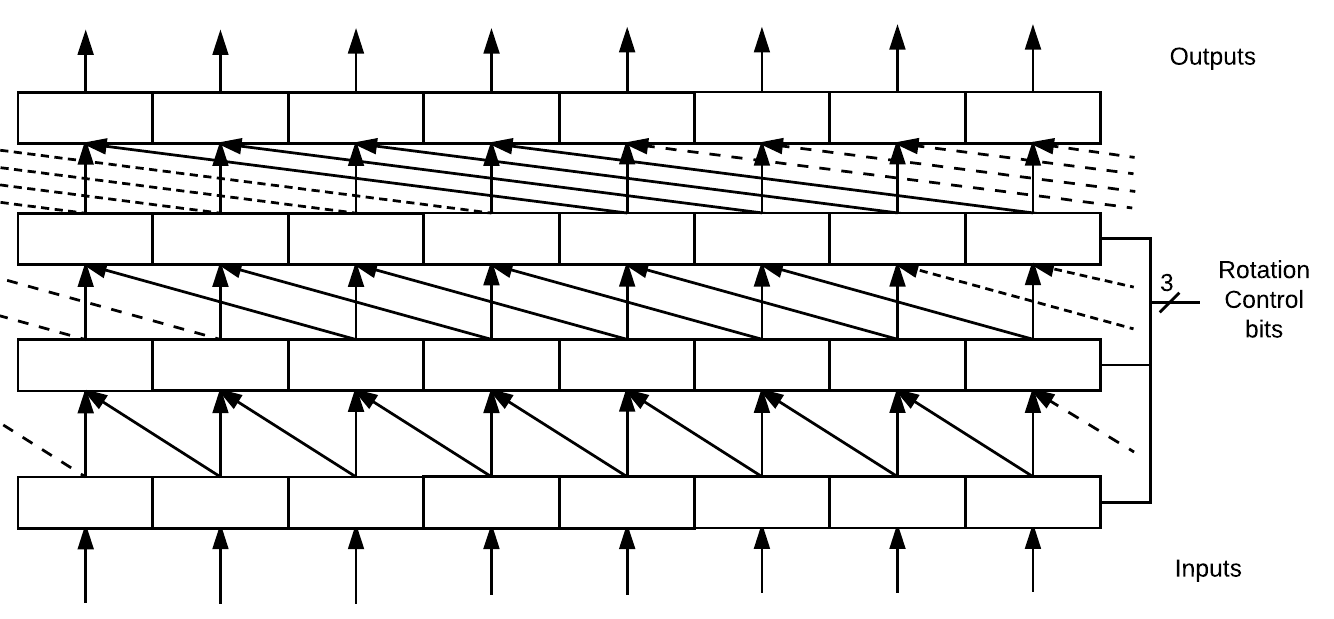}
\caption{An example data rotation unit for supporting eight ports.}
\label{fig:rotation_unit}
\end{figure}

The data rotation unit takes $N$ values of $W_{acc}$ bits each and left-rotates them in increments of $W_{acc}$ bits (rotating by $W_{acc}\times c$ bits in cycle $c$).
Figure~\ref{fig:rotation_unit} shows an example rotation unit with $N=8$ ports.
This unit, using a barrel shifter structure, passes data through $\log_2(N)$ levels of logic, where level $\ell$ is capable of rotating the word by the bit length of $2^\ell$ words.
Stage $\ell$ is controlled by bit $\ell$ of the binary encoding of the desired rotation amount, where logic-1 indicates that the stage should rotate.
Data rotation can either be performed in a single cycle or be pipelined, depending on the frequency requirements.

\subsection {Support for Burst Transfer}
\label{ssec:burst}

Support for burst data transfers is necessary to utilize the bandwidth available from the DRAM controller.

\subsubsection{Burst Transfer for Memory Read}

A request can generate a burst of line transfers to its port.
Therefore, the input buffer must be large enough to accommodate at least one burst per port.
In other words, the input buffer capacity must be at least $MaxBurstLen\times N$, with $N$ being the number of ports.
For each port, head and tail pointers are maintained to track its input buffer space.
In each cycle, only the lines at the head pointers participate in rotation.
A head pointer is incremented when the line it points to has finished transposition.
Tail pointers control where incoming memory lines are written.

\subsubsection{Burst Transfer for Memory Write}

The output buffer capacity must be at least $MaxBurstLen\times N$.
Similar to the case of memory read, head and tail pointers are used to keep track of the buffer space for each port.
Notably, for memory write, the request arbiter must monitor data coming from the write ports, and only issue requests for ports that have accumulated enough data in the output buffer to finish the write request.
This requirement also applies to the baseline interconnect.

\subsection {Interconnect Scalability}

The primary use of logic resources for our memory interconnect design comes from the data rotation unit.
There are $\log_2(N)$ layers of muxes in the data rotation unit, and each layer contains $N$ 2-to-1 muxes of width $W_{acc}$.
Overall, each layer contains $N\times W_{acc} = W_{line}$ 2-to-1 one-bit muxes, and all layers combined contain $W_{line} \times \log_2(N)$ many 2-to-1 one-bit muxes, which is a significant improvement over the baseline's cost of $W_{line} \times (N-1)$ muxes.

Furthermore, the Medusa interconnect consolidates the shallow and wide FIFOs of the baseline design into large buffers with deep and narrow banks, making them amenable to efficient storage in BRAM.

\subsection {Latency Overhead}
\label{ssec:latency_overhead}
Compared to the baseline, our transposition-based design has a \emph{constant} latency overhead of $W_{line}/W_{acc}$ cycles.
This happens because a memory line can only be consumed after it has been transposed.
For a typical case, $W_{line}/W_{acc} = 512/16 = 32$.
In the context of DNN accelerators, this latency overhead has a negligible impact on performance,
because DNN layer processors double buffer their inputs and perform perfect prefetch of data into the idle buffers.

Note that, even for burst transfers, the latency overhead of Medusa is still $W_{line}/W_{acc}$ cycles.
This is because as soon as the head of a burst arrives, transposition can start.

\subsection {Data Transfer Characteristics}
In the example in Figure~\ref{fig:detailed_read_transpose_example}, the buffer has data available for each port at the time when the transposition begins.
However, this is \emph{not} a requirement of the design.
The control logic starts transposition for a port without waiting for the other ports, and a port can join the transposition when transfers on the other ports are already in progress.
In other words, the transposition design does \emph{not} incur any interference among ports.

Overall, except for the constant latency overhead explained in Section~\ref{ssec:latency_overhead}, the data transfer characteristics of the Medusa interconnect are identical to that of the baseline.

\subsection {Handling Irregular Configurations}
\label{ssec:irregular_configuations}

Thus far, our discussion assumed a power-of-two number of read/write ports.
However, this is not a requirement.
When the number of ports is not a power of two, unused ports are either left unconnected (for output signals) or connected to suitable constant values (for input signals), and synthesis and place and route tools will perform suitable optimizations to remove unused logic.

\section{Evaluation}
\label{sec:evaluation}

We compare the Medusa transposition-based interconnect and the baseline interconnect by looking at their resource use, performance, and scalability.
Note that both interconnects use the same request arbitration logic, hence our evaluation focuses on the data transfer networks within the interconnects.

\subsection{Methodology}

We used Bluespec~\cite{bluespec} to implement both interconnects.
The implementations are highly parameterized to allow easy generation of various design points used in our experiments.
For all designs, we perform synthesis as well as place and route (P\&R) using Xilinx Vivado 2016.4, with the Virtex-7 690T FPGA as the target device.

An important aspect of our evaluation is finding the post-P\&R peak frequencies of different data transfer networks.
However, doing P\&R for the data transfer networks alone will not yield representative results.
This is because within a DNN accelerator, layer processors consume the most resources and thus will have a significant impact on the results of place and route.
Since the DNN accelerator's performance is what ultimately matters, ignoring the P\&R impact from layer processors is unreasonable.
As such, when running synthesis and P\&R for a memory interconnect, we also include into the design a convolutional layer processor, which uses all the narrow read/write ports of the interconnect.
At a high level, a convolutional layer processor consists of vector dot-product units, input feature map buffers, output feature maps buffers, and weight buffers~\cite{chen2014diannao, zhang2015optimizing, shen2017multiclp}.
In our case, the number of vector dot-product units is set differently for different experiments.
Each vector dot-product unit is 32-wide and operates on vectors of 16-bit fixed point values.
Correspondingly, the narrow read/write ports of an interconnect are 16-bits wide.
Each vector dot-product unit uses 32 DSP slices to implement its 32 multipliers.
Input feature map buffers are 2260 deep, output feature map buffers are 1792 deep, and weight buffers are 9 deep.
These buffer depths are chosen to be suitable for VGGNet~\cite{simonyan2014very} and similar CNNs, and result in BRAM use comparable to layer processors in existing works~\cite{zhang2015optimizing,shen2017multiclp}.

To compare the resource use of different designs, we look at the post-P\&R consumption of the four main types of FPGA resources: LUTs (look up tables), FFs (flip-flops), 18Kbit BRAMs (block RAMs), and DSPs (DSP slices).

To compare the performance of different designs, we find the post-P\&R peak obtainable frequency of each design (searching in steps of 25MHz).
We used Vivado's default synthesis strategy with retiming turned on.
For place and route, we used the ``performance explore plus post-route optimization'' strategy.

For the scalability evaluation, we vary the layer processor and interconnect size, and observe the changes in peak frequency.

To ease the implementation process, we excluded the memory controller and PCIe controller from our setup and replaced them with stubs.
The exclusion of these two components gives \emph{equal benefit} to the baseline designs and Medusa transposition-based designs in terms of their area consumed and their ability to reach higher frequencies.
Importantly, because these two components only use a small fraction of the resource of a Virtex-7 690T and run in their own clock domains, the frequency benefit from their exclusion is minor, and will not affect the conclusions of our experiments.

\subsection{Baseline Validation}

Even though off-the-shelf IP cores can be used to implement the baseline read and write interconnects, they often have limitations that are insufficient for our use. For example, the Xilinx AXI4-Stream Interconnect only supports up to 16 ports, but we often require more (e.g., to consume a 512-bit DDR3 interface, we need 32 ports of 16 bits each). 

To validate that our baseline implementation is resource-efficient, we compare the post-synthesis resource use of our baseline data transfer networks with equivalent networks built from Xilinx AXI4-Stream IP cores.
Xilinx AXI4-Stream IP cores are used because they do not include the overhead of AXI request arbitration, much like the data transfer networks that we aim to evaluate.
To enable this comparison, we choose a configuration that fits within the limits of Xilinx AXI4-Stream IP cores---specifically using a 256-bit wide memory interface, multiplexed to 16 16-bit ports. We set the FIFO depth to 32 words.
As Table~\ref{tab:baseline_validation} shows, our baselines have significantly lower cost than their Xilinx AXI4-Stream IP-based counterparts.
As such, our baseline designs represent fair reference points with which to compare our Medusa transposition-based designs.

\begin{table}
\caption{Baseline data transfer networks vs. AXI4-Stream networks (1$\times$256-bit port to 16$\times$16-bit ports. No DSPs or BRAMs are used).}
\label{tab:baseline_validation}
\centering
\begin{tabular}{@{}lrrrr@{}}
\toprule
      & Base (Read) & AXIS (Read) & Base (Write) & AXIS (Write)\\
\midrule
LUT   & 5,313     & 11,562    & 6,810     & 9,170\\
      & (1.2\%)   & (2.7\%)   & (1.6\%)   & (2.1\%) \\
\midrule
FF    & 5,404     & 27,173    & 9,023     & 26,554\\
      & (0.6\%)   & (3.1\%)   & (1.0\%)   & (3.1\%) \\
\bottomrule
\end{tabular}
\end{table}

Note that the data transfer networks in Table~\ref{tab:baseline_validation} are relatively small, and do not exhibit the resource consumption problems we will see when scaling to larger networks.
The following sections investigate larger designs to show this problem.

\subsection{Hardware Resource Usage}
\label{ssec:eval_resource_use}

To evaluate the hardware resources required by Medusa, we thoroughly evaluate a representative design point.
The design point we use include a layer processor with 64 vector dot-product units.
We evaluate this layer processor coupled with a memory interconnect, built using either the baseline or the Medusa transposition-based data transfer networks.
Each memory interconnect multiplexes a 512-bit memory interface to 32 16-bit read ports and 32 16-bit write ports; all ports are used by the layer processor.
For each read/write port, the maximum burst that the memory interconnect can buffer is 32$\times$512-bits.
Overall, this setup represents a scenario where the external memory is a single channel 800MHz DDR3, which is common on FPGA boards.
In such systems, the memory controller runs in its own clock domain at 200MHz, and exposes a 512-bit interface to the rest of the FPGA.
The amount of DSP slices and BRAM slices used by the layer processor in this setup is representative of existing works~\cite{zhang2015optimizing, shen2017multiclp, shen2017escher}.

Table~\ref{tab:resouce_use} shows the resource breakdown of the two designs.
For each design, we present the resource use of the whole design, the read data-transfer network, and the write data-transfer network in isolation.
The percentages show resource use relative to the capacity of a Virtex-7 690T.

\begin{table}
    \caption{Medusa vs. Baseline (FPGA Resource Use).}
    \label{tab:resouce_use}
    \centering
    \begin{tabular}{@{}llrrrr@{}}
\toprule
        &              & LUT & FF & BRAM-18K & DSP\\
\midrule
        & Read Network & 18,168 &  19,210 & 0          &     0\\
        &              & (4.2\%) & (2.2\%) & (0\%)      &  (0\%)\\
\cmidrule{2-6}
Baseline & Write Network   & 26,810    &  35,451  & 0          &     0\\
         &                 & (6.2\%)   & (4.1\%)  & (0\%)      &  (0\%)\\
\cmidrule{2-6}
         & Total       & 198,887   & 240,449  & 726        & 2,048\\
         &             & (45.9\%)  & (27.8\%) & (24.7\%)   & (56.9\%)\\
\midrule
        & Read Network   & 4,733  & 4,759  & 32         &       0\\
        &                & (1.1\%)  & (0.6\%)& (1.1\%)  &   (0\%)\\
 \cmidrule{2-6}
Medusa  & Write Network  & 4,777  & 4,325 & 32         &       0\\
        &                & (1.1\%)   & (0.5\%) & (1.1\%)    &   (0\%)\\
\cmidrule{2-6}
        & Total        & 156,409  & 195,158   & 790        & 2,048\\
        &              & (36.1\%)  & (22.5\%) & (26.9\%)   & (56.9\%)\\
\bottomrule
\end{tabular}
\end{table}

First, we focus on the data transfer networks in isolation.
For memory read, compared to the baseline, the Medusa transposition-based network reduces LUT use by 3.84x and FF use by 4.04x, at a cost of 32 BRAMs.
For memory write, the Medusa transposition-based network reduces LUT use by 5.61x and FF use by 8.20x, also at a cost of 32 BRAMs.
Combined, the Medusa networks achieve 4.73x LUT and 6.02x FF savings, at a minor BRAM cost.

We next consider the entire design, including the layer processor and memory interconnect (Total).
The baseline uses 1.27x more LUTs and 1.23x more FFs than the Medusa transposition-based design, whereas the transposition-based design uses 1.09x more BRAM.
This shows that the LUT and FF savings achieved by the Medusa data transfer networks are significant even in the context of a resource-heavy layer processor.
In the baseline design, the combined read and write data-transfer networks account for 22.6\% of the total LUT use and 22.7\% of the total FF use of the accelerator.
The Medusa transposition-based design reduced these to 6.1\% and 4.7\%, respectively.

The Medusa network's efficiency stems from its lower logic complexity and its ability to make efficient use of BRAMs, saving LUTs, FFs, and routing resources.
The Medusa design uses a total of 64 BRAMs to efficiently buffer data.
In contrast, if the baseline design were to use BRAMs in its data-transfer networks, 960 BRAMs would be needed, making it a poor trade-off with respect to the savings in FFs and LUTs.
This is because each 18-Kbit BRAM is 36 bits wide, and each 32x512-bit FIFO would consume 15 BRAMs, requiring a total of 960 BRAMs for 32 memory-read FIFOs and 32 memory-write FIFOs.

\subsection{Performance and Scalability}

Besides saving logic resources, our Medusa transposition-based design also critically saves on routing resources.
The compounded savings of logic and routing can lead to significant improvements in an accelerator's performance and scalability.
In this section, we evaluate this important quality of our Medusa memory interconnect by scaling the size of both the accelerator and the interconnect, and observing how the baseline and Medusa change the reachable clock frequency. 

Our experiment starts with a small layer processor with 16 vector dot-product units.
The memory interconnect multiplexes a 128-bit memory interface to 8 16-bit read ports and 8 16-bit write ports.
We perform place and route to find the peak obtainable frequency of the accelerator, with a search step of 25MHz, for both the baseline and for the Medusa-based design.
From this point, we go through several steps where we scale up the accelerator's size and number of memory ports, and we repeat the peak frequency search.
At each step, we increase the number of vector dot-products units by 8, the number of read ports by 4, and the number of write ports by 4.
The width of the memory interface is always set to a power of two, and is chosen to be wide enough to accommodate all the read ports and write ports.
For example, for $(8, 16]$ 16-bit read ports, a 256-bit interface is needed, and for $(16, 32]$ read ports, 512-bits are needed.

\begin{figure}
\centering
\includegraphics[width=\columnwidth]{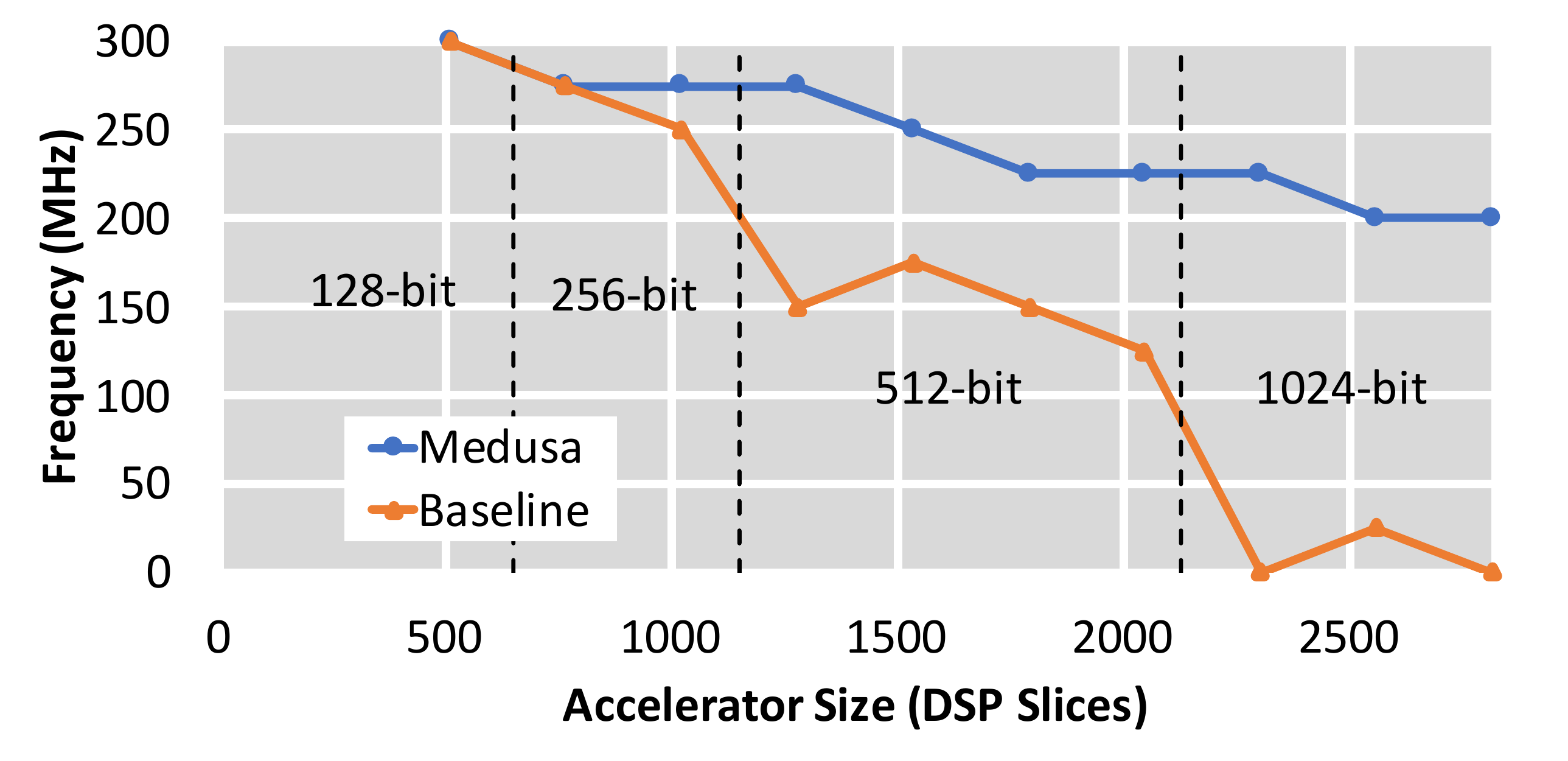}
\caption{Change in peak frequency as the accelerator scales.}
\label{fig:scaling}
\end{figure}

Figure~\ref{fig:scaling} shows the result of our experiment. The x-axis shows the size of the accelerator, measured in DSP slices (equal to the number of vector dot-product units times 32).
The y-axis indicates the maximum reachable frequency for that point. Separate lines are used to indicate the accelerator implemented using the Medusa interconnect and the baseline interconnect; keep in mind that the only difference between the two lines is the interconnect.
Points at 0MHz indicate that Vivado was not able meet timing at 25MHz.
Vertical dashed lines are used to partition this figure into four regions, where each region corresponds to a memory interface width (128-bit through 1024-bit).

From Figure~\ref{fig:scaling}, we can see that starting from the point with 1024 DSPs, Medusa designs always outperform baseline designs.
Furthermore, as the size of the accelerator increases, the performance difference between Medusa and the baseline increases, demonstrating the improved scalability of Medusa designs.

Besides the general trend, there are several other interesting things to note in Figure~\ref{fig:scaling}.
First, within the 512-bit memory interface region, which is the region that represents the popular configuration of ``FPGA + single channel DDR3,''
Medusa outperforms the baseline by up to 1.8x (the designs with 1280 DSPs and 2048 DSPs).
Second, within the 1024-bit memory interface region, which represents platforms with higher memory bandwidth, the baseline is barely usable, with some points failing to make timing even at 50MHz or lower.
Nonetheless, the Medusa designs with the same large accelerators can keep running at 200 to 225MHz in this region.
Third, we note that although Medusa designs are most efficient when the number of memory ports is a power of two, most of the designs included in our experiment have a non-two-power port count.
In spite of this, Medusa still shows a clear benefit over the baseline.
Lastly, we note that the 2048-DSP points in Figure~\ref{fig:scaling} correspond to the designs whose resource use metrics were evaluated in Table~\ref{tab:resouce_use}.
This demonstrates that the Medusa designs simultaneously achieve both resource savings and performance improvements.

\section{Related Work}
\label{sec:related}

Our work focuses on providing an efficient memory interconnect for DNN accelerators that require access to DRAM through many narrow read and write ports.
Some designs~\cite{caffeine2016zhang, qiu2016going} avoid the need for such an interconnect by altering the layout of data in DRAM.
The main drawback of this approach is its constraint on the data flow inside layer processors.
Specifically, the output layout of one layer must be compatible with the input layout of the next, limiting how a layer processor can perform its computation, which can lead to underutilization of compute units~\cite{shen2017multiclp}.

Other designs~\cite{sharma2016dnnweaver,chen2016eyeriss_isca} avoid the width mismatch problem by using narrow memory controller buses.
When scaled up, these designs will either be bottlenecked by DRAM bandwidth, or face the interconnect conundrum which is addressed by Medusa.
In particular, \cite{sharma2016dnnweaver} uses a multi-cast network for distributing read data to the accelerators, and a mux-based design for writing.
Data being read from the memory interface are broadcast to all ports with an ID attached.
A port can decide whether to accept the data by checking the ID.
From a resource use point of view, the multi-cast network is essentially the read data network in our baseline; this and the mux-based write interconnect will have similar scalability problems as the baseline designs considered in this work.

Yet other designs~\cite{umuroglu2017finn,brainwave} avoid the need for an advanced memory interconnect by storing DNN model parameters as well as intermediate data on chip, so as to reduce bandwidth use and simplify the interaction between the accelerator and DRAM.
For such designs, the supported DNN size is limited by on-chip storage size.

\section{Conclusions}
\label{sec:conclusions}

This paper presented a resource efficient and high-performance memory interconnect for connecting many-port DNN accelerators to wide DRAM controller interfaces.
We analyzed and experimented with commonly-used mux/demux-based interconnects, and concluded that they were over-provisioned and had serious scalability limitations.

To address this problem, we tailored our design to the needs of DNN accelerators and used a transposition unit to implement memory bandwidth partitioning.
Our design has lower logic complexity and can efficiently use BRAMs to reduce LUTRAM use.
Experiments showed that, compared to the baseline design, our Medusa design reduced LUT and FF usage by 4.7x and 6.0x respectively, and improved peak frequency by 1.8x.

% use section* for acknowledgement
\section*{Acknowledgment}

This material is based on work supported by the National Science Foundation (NSF) under Grant Nos. 1533739 and 1453460. The experiments were conducted with equipment purchased through NSF CISE Research Infrastructure Grant No. 1405641.

\bibliographystyle{IEEEtranBST2/IEEEtran}
\bibliography{refs}

\end{document}